\documentclass[11pt,a4paper]{article}

\usepackage{amssymb}
\usepackage{graphicx}
\usepackage{bm}

\usepackage{mathrsfs}
\usepackage{cite}

\begin{document}
	
\title{A condition for the reduction of couplings in the $P = \frac{1}{3}Q$ supersymmetric theories}

\author{M.D.Kuzmichev\,${}^{ab}$, K.V.Stepanyantz\,${}^{abc}$ $\vphantom{\Big(}$
\medskip\\
{\small{\em Moscow State University, Faculty of Physics,}}\\
${}^a${\small{\em Department of Theoretical Physics,}}\\
${}^b${\small{\em Department of Quantum Theory and High Energy Physics,}}\\
{\small{\em 119991, Moscow, Russia}}\\
$^c$ {\small{\em Bogoliubov Laboratory of Theoretical Physics, JINR,}}\\
{\small{\em 141980 Dubna, Moscow region, Russia.}}\\}

\maketitle
	
\begin{abstract}
We demonstrate that in the $P=\frac{1}{3}Q$ supersymmetric theories the renormalization group invariance of the ratio $\lambda^{ijk}/e$ (of the Yukawa couplings to the gauge coupling) is equivalent to a simple relation between the anomalous dimensions of the quantum gauge superfield, of the Faddeev--Popov ghosts, and of the matter superfields, which should be valid in each order of the perturbation theory. In the one- and two-loop approximations it is verified explicitly. Presumably, in higher orders this relation can be satisfied for the planar supergraphs under a certain renormalization prescription. Assuming that it is valid we rewrite the exact equation for the (corresponding contribution to the) anomalous dimension of the matter superfields in the theories under consideration in a different (but equivalent) form.
\end{abstract}
	
	\unitlength=1cm
	
\section{Introduction}
\hspace*{\parindent}

In ${\cal N}=2$ supersymmetric gauge theories the Yukawa couplings are proportional to the gauge coupling constant. If a regularization and a renormalization prescription do not break ${\cal N}=2$ supersymmetry, then this relation remains valid at the quantum level. The most natural way to achieve this is to use the harmonic superspace \cite{Galperin:1984av,Galperin:2001seg,Buchbinder:2001wy}, which makes ${\cal N}=2$ supersymmetry manifest at all steps of calculating quantum corrections. However, in the case of using the ${\cal N}=1$ superspace formulation of ${\cal N}=2$ supersymmetric gauge theories, the relation between the Yukawa and gauge couplings can in general be broken if a regularization and a renormalization prescription are not ${\cal N}=2$ supersymmetric. Nevertheless, a renormalization prescription for which the ratio $\lambda^{ijk}/e$ (of the Yukawa couplings to the gauge coupling) remains renormalization group (RG) invariant exists in these theories \cite{Aleshin:2023ior}. It is interesting to reveal if it is possible to construct other ${\cal N}=1$ theories in which this ratio is RG invariant. (Certainly, ${\cal N}=2$ supersymmetric theories is a particular case of ${\cal N}=1$ supersymmetric theories.) It was suggested \cite{Jack:1996qq} that in certain ${\cal N}=1$ supersymmetric theories the ratio $\lambda^{ijk}/e$ can really be RG invariant.\footnote{Note that this implies that a number of couplings can be reduced. The recent review devoted to the reduction of couplings
\cite{Zimmermann:1984sx,Oehme:1984yy,Kubo:1985up} and the corresponding references can be found in \cite{Heinemeyer:2019vbc}.} According to \cite{Jack:1996qq}, this could possibly occur in ${\cal N}=1$ supersymmetric theories which satisfy the so-called $P=\frac{1}{3}Q$ constraint \cite{Jack:1995gm},

\begin{equation}\label{PQ3_Constraint}
\lambda^*_{imn}\lambda^{jmn} - 4\pi\alpha C(R)_i{}^j = \frac{2\pi\alpha}{3} Q \delta_i^j,
\end{equation}

\noindent
where

\begin{equation}
Q \equiv T(R) - 3C_2
\end{equation}

\noindent
and $\alpha \equiv e^2/4\pi$. Here we assume that the gauge group $G$ (of the dimension $r$) is simple and the generators of its fundamental representation are normalized by the condition $\mbox{tr}(t^A t^B)=\delta^{AB}/2$. The matter superfields belong to the representation $R$ of the group $G$, which in general can be reducible. The generators in this representation are denoted by $T^A$, and the group Casimirs are defined as

\begin{equation}\label{Casimirs}
\mbox{tr}\,(T^A T^B) = T(R)\,\delta^{AB};\qquad (T^A T^A)_i{}^j = C(R)_i{}^j;\qquad C_2 = T(Adj).
\end{equation}

\noindent
For $Q=0$ the $P=\frac{1}{3}Q$ theories are reduced to the one-loop finite ${\cal N}=1$ supersymmetric theories \cite{Parkes:1984dh,Kazakov:1986bs,Ermushev:1986cu,Lucchesi:1987he,Lucchesi:1987ef}, see \cite{Heinemeyer:2019vbc} for a recent review.
	
If an ${\cal N}=1$ supersymmetric gauge theory satisfies the constraint (\ref{PQ3_Constraint}), then in the one-loop approximation

\begin{equation}\label{RG_Invariance}
\frac{d}{d\ln\mu}\Big(\frac{\lambda^{ijk}}{e}\Big) = 0,
\end{equation}

\noindent
where $\mu$ is a renormalization point, see \cite{Jack:1996qq}. It appears that in the two-loop approximation this equation is also valid in the $\overline{\mbox{DR}}$-scheme (when a theory is regularized by dimensional reduction \cite{Siegel:1979wq}, and divergences are removed by modified minimal subtraction \cite{Bardeen:1978yd}). However, in the three-loop approximation Eq. (\ref{RG_Invariance}) is not satisfied, and it is even impossible to obtain it by a special tuning of a renormalization scheme \cite{Jack:1996qq}. Presumably, this implies that some more constraints should be imposed on the theory or this equation is valid in a certain approximation. To make a step towards finding the conditions under which Eq. (\ref{RG_Invariance}) is valid, in this paper we rewrite it in the form of a relation between the renormalization group functions (RGFs) of quantum superfields, namely, of the quantum gauge superfield, of the Faddeev--Popov ghosts, and of the matter superfields. In what follows we will see that the corresponding anomalous dimensions should be related to each other in each order of the perturbation theory, see Eq. (\ref{Gamma_Relation}) below.

The paper is organized as follows. In Sect. \ref{Section_RG_Invariance} from Eq. (\ref{RG_Invariance}) we derive the relation between the anomalous dimensions of the quantum superfields for the $P=\frac{1}{3}Q$ theories and verify that it is valid in the one-loop approximation. Using this relation in Sect. \ref{Section_Exact_Equations} we rewrite the exact expression for the anomalous dimension of the matter superfields in a new form, which can presumably be used for its derivation. The validity of the relation between the anomalous dimensions of quantum superfields in higher orders is discussed in Sect. \ref{Section_Higher_Orders}. Conclusion contains a brief summary of the results.

\section{Reduction of couplings and the anomalous dimensions of quantum superfields}
\hspace{\parindent}\label{Section_RG_Invariance}

Due to the nonrenormalization of the superpotential \cite{Grisaru:1979wc} it is possible to choose a renormalization prescription for which

\begin{equation}\label{Lambda_Renormalization}
\lambda^{ijk} = \lambda_0^{mnp} \big(\sqrt{Z_\phi}\big)_m{}^i \big(\sqrt{Z_\phi}\big)_n{}^j \big(\sqrt{Z_\phi}\big)_p{}^k,
\end{equation}

\noindent
where $\lambda_0^{ijk}$ and $\lambda^{ijk}$ are the bare and renormalized Yukawa couplings, respectively, and $(Z_\phi)_i{}^j$ denotes the renormalization constants for the chiral matter superfields $\phi_i$. On the other hand, it is possible to relate the renormalization constants using the nonrenormalization of the triple gauge-ghost vertices derived in \cite{Stepanyantz:2016gtk} with the help of the Slavnov--Taylor identities \cite{Taylor:1971ff,Slavnov:1972fg} and the rules for calculating supergraphs. According to this theorem, the vertices with two external ghost legs and one external leg corresponding to the {\it quantum} gauge superfield are finite in all orders. Consequently, the renormalization constants can be chosen in such a way that

\begin{equation}\label{VCC_Finiteness}
Z_{\alpha}^{-1/2} Z_c Z_V = 1.
\end{equation}

\noindent
Note that in our notation they are defined by the equations

\begin{equation}
\frac{1}{\alpha_0} = \frac{Z_\alpha}{\alpha};\qquad V = Z_V Z_\alpha^{-1/2} V_R; \qquad \bar c\, c = Z_c Z_\alpha^{-1} \bar c_R c_R; \qquad \phi_i = \big(\sqrt{Z_\phi}\big)_i{}^j \big(\phi_{R}\big)_j,
\end{equation}

\noindent
where $V$ is the quantum gauge superfield, while $c$ and $\bar c$ are the (chiral) ghost and antighost superfields, respectively. The bare coupling constant is denoted by $\alpha_0$, and the subscript $R$ indicates the renormalized superfields.

Substituting Eqs. (\ref{Lambda_Renormalization}) and (\ref{VCC_Finiteness}) into Eq. (\ref{RG_Invariance}) we obtain the condition

\begin{eqnarray}
&& 0 = \frac{\lambda_0^{mnp}}{e_0} \frac{d}{d\ln\mu}\Big(\frac{1}{\sqrt{Z_\alpha}} \big(\sqrt{Z_\phi}\big)_m{}^i \big(\sqrt{Z_\phi}\big)_n{}^j \big(\sqrt{Z_\phi}\big)_p{}^k\Big)
\nonumber\\
&&\qquad\qquad\qquad\qquad\qquad
=  \frac{\lambda_0^{mnp}}{e_0} \frac{d}{d\ln\mu}\Big(Z_c^{-1} Z_V^{-1} \big(\sqrt{Z_\phi}\big)_m{}^i \big(\sqrt{Z_\phi}\big)_n{}^j \big(\sqrt{Z_\phi}\big)_p{}^k\Big),\qquad
\end{eqnarray}

\noindent
which is evidently satisfied if

\begin{equation}
\frac{d}{d\ln\mu}\Big((Z_c)^{-2/3} (Z_V)^{-2/3} \big(Z_\phi\big)_i{}^j\Big) = 0.
\end{equation}

\noindent
This implies that the anomalous dimensions of quantum superfields satisfy the equation

\begin{equation}\label{Gamma_Relation}
2\big(\gamma_c + \gamma_V\big)\, \delta_i^j = 3\big(\gamma_\phi\big)_i{}^j.
\end{equation}

\noindent
Obviously, this relation should be valid in each order of the perturbation theory. The one-loop expressions for $\gamma_c$ and $\gamma_V$ can be found in, e.g., \cite{Aleshin:2016yvj},

\begin{eqnarray}
&& \gamma_c^{(1)} = -\frac{\alpha C_2(1-\xi)}{6\pi};\qquad \gamma_V^{(1)} = \frac{\alpha C_2(1-\xi)}{6\pi} + \frac{Q\alpha}{4\pi};\qquad \nonumber\\
&& \big(\gamma_\phi^{(1)}\big)_i{}^j = -\frac{\alpha}{\pi} C(R)_i{}^j + \frac{1}{4\pi^2}\lambda^*_{imn}\lambda^{jmn} = \frac{Q\alpha}{6\pi}\delta_i^j,
\end{eqnarray}

\noindent
where $\xi$ is the gauge parameter (such that the coefficient in the gauge fixing term is proportional to $1/\xi$), and in the last equality we took into account the $P=\frac{1}{3}Q$ condition (\ref{PQ3_Constraint}). From these expressions we see that the relation (\ref{Gamma_Relation}) is really satisfied in the one-loop approximation.

\section{Exact equations for $P=\frac{1}{3}Q$ theories}
\hspace*{\parindent}\label{Section_Exact_Equations}

According to \cite{Jack:1996qq}, Eq. (\ref{RG_Invariance}) leads to the exact equation for the anomalous dimension of the matter superfields. For completeness, here we briefly describe the derivation of this equation from Eq. (\ref{Gamma_Relation}). Due to Eq. (\ref{VCC_Finiteness}) the $\beta$-function is related to the anomalous dimensions of the Faddeev--Popov ghosts and of the quantum gauge superfield by the equation \cite{Stepanyantz:2016gtk}

\begin{equation}\label{Beta_VCC}
\beta = 2\alpha(\gamma_c + \gamma_V).
\end{equation}

\noindent
From Eqs. (\ref{Gamma_Relation}) and (\ref{Beta_VCC}) we obtain the relation (first derived in \cite{Jack:1996qq} directly from Eq. (\ref{RG_Invariance}))

\begin{equation}\label{Beta_Gamma_Relation}
\big(\gamma_\phi\big)_i{}^j = \frac{\beta}{3\alpha}\,\delta_i^j.
\end{equation}

Next, it is necessary to involve the Novikov, Shifman, Vainshtein, and Zakharov (NSVZ) equation \cite{Novikov:1983uc,Jones:1983ip,Novikov:1985rd,Shifman:1986zi}, which relates the $\beta$-function to the anomalous dimension of the chiral matter superfields in ${\cal N}=1$ supersymmetric theories,

\begin{equation}\label{NSVZ_Original}
\beta(\alpha, \lambda) = \frac{\alpha^2 \big( Q - C(R)_i{}^j (\gamma_\phi)_j{}^i(\alpha, \lambda)/r\big)}{2\pi(1-C_2\alpha/2\pi)}.
\end{equation}

\noindent
Its perturbative derivation has been done in \cite{Stepanyantz:2016gtk,Stepanyantz:2019ihw,Stepanyantz:2020uke} (see also \cite{Stepanyantz:2019lfm}). This derivation allowed to construct an all-loop prescription giving some NSVZ schemes, in which Eq. (\ref{NSVZ_Original}) is valid in all orders. Namely, it is necessary to regularize a theory by the higher covariant derivative (HD) method \cite{Slavnov:1971aw,Slavnov:1972sq,Slavnov:1977zf} in the superfield version \cite{Krivoshchekov:1978xg,West:1985jx} (see also \cite{Aleshin:2016yvj,Kazantsev:2017fdc}) and use minimal subtractions of logarithms (MSL) \cite{Kataev:2013eta,Shakhmanov:2017wji} for removing divergences. This prescription can supplement various versions of this regularization and implies that only powers of $\ln\Lambda/\mu$ are included into the renormalization constants, where $\Lambda$ is the regularization parameter with the dimension of mass. The resulting HD+MSL schemes are NSVZ schemes in all orders of the perturbation theory.

From Eqs. (\ref{Beta_Gamma_Relation}) and (\ref{NSVZ_Original}) it is possible to derive the exact expressions \cite{Jack:1996qq} for the $\beta$-function and for the anomalous dimension of the matter superfields for the $P=\frac{1}{3}Q$ theories (which should be valid if the assumption (\ref{RG_Invariance}) is correct),

\begin{equation}\label{Beta_Gamma_Exact}
\beta(\alpha) = \frac{\alpha^2 Q}{2\pi (1+\alpha Q/6\pi)};\qquad (\gamma_\phi)_i{}^j(\alpha) = \frac{\alpha Q}{6\pi (1+\alpha Q/6\pi)}\delta_i^j \equiv \gamma_\phi \delta_i^j.
\end{equation}

\noindent
Note that they depend only on the gauge coupling constant and are independent of the Yukawa couplings. We see that the expressions for both the anomalous dimension and the $\beta$-function have the form of the geometric series. This structure is very similar to the expression for the exact $\beta$-function of the pure ${\cal N}=1$ supersymmetric Yang--Mills theory \cite{Jones:1983ip}. That is why it is reasonable to suggest that a possible derivation of the exact equations (\ref{Beta_Gamma_Exact}) could be made in a similar way. However, the perturbative derivation of the exact NSVZ $\beta$-function does not originally produce it in the form (\ref{NSVZ_Original}), see \cite{Stepanyantz:2019lfm,Stepanyantz:2020uke}. In fact, it gives a different form of the NSVZ equation proposed in \cite{Stepanyantz:2016gtk}, which is obtained from Eqs. (\ref{Beta_VCC}) and (\ref{NSVZ_Original}),

\begin{equation}\label{NSVZ_New}
\beta(\alpha, \lambda) = \frac{\alpha^2}{2\pi}\Big( Q + 2C_2\gamma_c(\alpha, \lambda) + 2C_2 \gamma_V(\alpha, \lambda) - C(R)_i^{\ j} (\gamma_\phi)_j^{\ i}(\alpha, \lambda)/r \Big).
\end{equation}

\noindent
This form does not contain a coupling dependent denominator and, therefore, relates the $\beta$-function in a certain loop to the anomalous dimensions of quantum superfields in the previous loop. Therefore, one may suggest that a possible derivation of the exact expression for the anomalous dimension in Eq. (\ref{Beta_Gamma_Exact}) could be done in a similar way. That is why it is reasonable to rewrite the second equation in (\ref{Beta_Gamma_Exact}) in a similar form (without the coupling dependent denominator). For this purpose we present it in the form

\begin{equation}\label{Gamma_Recursion}
\gamma_\phi(\alpha) = \frac{\alpha Q}{6\pi}\Big(1 - \gamma_\phi(\alpha) \Big),
\end{equation}

\noindent
which relates the anomalous dimension of the matter superfields in $L+1$ and $L$ loops. With the help of Eq. (\ref{Gamma_Relation}) this equation can equivalently be rewritten as the relation

\begin{equation}\label{Gamma_With_Ghosts}
\gamma_\phi(\alpha) = \frac{\alpha Q}{6\pi}\Big(1 - \frac{2}{3}(\gamma_c + \gamma_V)\Big)
\end{equation}

\noindent
between the anomalous dimension of the matter superfields in a certain loop and the anomalous dimensions of the Faddeev--Popov ghosts and of the quantum gauge superfield in the previous loop. Moreover, it is also possible to write a more general expression

\begin{equation}\label{Gamma_With_X}
\gamma_\phi(\alpha) = \frac{\alpha Q}{6\pi}\Big(1 - \frac{2}{3}x\left(\gamma_c(\alpha) + \gamma_V(\alpha)\right) - (1-x)\gamma_\phi(\alpha)\Big),
\end{equation}

\noindent
where $x$ is an arbitrary real number. For $x=0$ and $x=1$ it gives Eq. (\ref{Gamma_Recursion}) and Eq. (\ref{Gamma_With_Ghosts}), respectively. At present, we are not able to pick out such a value of $x$ for which Eq. (\ref{Gamma_With_X}) can be derived by perturbative methods (if possible).

\section{Higher orders}
\hspace*{\parindent}\label{Section_Higher_Orders}

In the end of Sect. \ref{Section_RG_Invariance} we demonstrated that Eq. (\ref{Gamma_Relation}) is really valid in the one-loop approximation. Here we discuss if this equation is valid in higher orders.

The two-loop anomalous dimensions of the Faddeev--Popov ghosts and of the chiral matter superfields were calculated for theories regularized by higher covariant derivatives in \cite{Kazantsev:2018kjx} and \cite{Kazantsev:2020kfl}, respectively. However, the two-loop expression for the anomalous dimension $\gamma_V$ has not yet been calculated directly. However, it is possible to construct it using the results of
\cite{Kuzmichev:2021yjo,Kuzmichev:2021lqa}, where it was demonstrated that the triple gauge-ghost vertices for ${\cal N}=1$ supersymmetric theories regularized by higher covariant derivatives \cite{Slavnov:1971aw,Slavnov:1972sq,Slavnov:1977zf} in the superfield version \cite{Krivoshchekov:1978xg,West:1985jx} are finite in the two-loop approximation.\footnote{Actually, this calculation has been done using the version of the higher derivative regularization proposed in \cite{Aleshin:2016yvj,Kazantsev:2017fdc}.} This implies that in this order Eq. (\ref{Beta_VCC}) is valid for RGFs defined in terms of the bare couplings, and the anomalous dimension of the quantum gauge superfield defined in terms of the bare couplings can be found from it. Then the standard (i.e., defined in terms of the renormalized couplings) anomalous dimension $\gamma_V$ can be calculated using the technique described in, e.g., \cite{Shirokov:2022jyd}. Certainly, this anomalous dimension (as well as all RGFs defined in terms of the renormalized couplings) depends on finite constants which specify a renormalization prescription. In the lowest approximation these constants are defined by the equations

\begin{eqnarray}
&& \alpha_0 = \alpha - \frac{\alpha^2}{2\pi}\bigg[3 C_2 \Big(\ln\frac{\Lambda}{\mu} + b_{11}\Big) - T(R) \Big(\ln\frac{\Lambda}{\mu} + b_{12}\Big)\bigg] + \ldots;\nonumber\\
&& \alpha_0\xi_0 = \alpha\xi + \frac{\alpha^2 C_2}{3\pi}\Big(\xi(\xi-1)\ln\frac{\Lambda}{\mu}  + x_1\Big)+\ldots;\nonumber\\
&& y_0 = y + \frac{\alpha}{90\pi}\Big((2+3\xi)\ln\frac{\Lambda}{\mu} + k_1\Big)+\ldots;\nonumber\\
&& \big(Z_\phi\big)_i{}^j = \delta_i^j + \frac{\alpha}{\pi} C(R)_i{}^j \Big(\ln\frac{\Lambda}{\mu} + g_{11}\Big) - \frac{1}{4\pi^2} \lambda^*_{imn}\lambda^{jmn} \Big(\ln\frac{\Lambda}{\mu} + g_{12}\Big)+\ldots;\nonumber\\
&& \ln Z_c = -\frac{\alpha C_2 (\xi-1)}{6\pi} \ln\frac{\Lambda}{\mu} + \frac{\alpha C_2}{\pi} h_1 + \ldots;\nonumber\\
&& \ln Z_V = \frac{\alpha}{\pi}\bigg\{C_2 \Big(\frac{3}{4}\ln\frac{\Lambda}{\mu} + \frac{1}{6}(\xi-1)\ln\frac{\Lambda}{\mu} +\frac{3}{4} v_1\Big) - \frac{1}{4} T(R) \Big(\ln\frac{\Lambda}{\mu} + v_2\Big)\bigg\}+\ldots,
\qquad
\end{eqnarray}

\noindent
where $y$ is the (first) parameter of the nonlinear renormalization of the quantum gauge superfield \cite{Piguet:1981fb,Piguet:1981hh,Tyutin:1983rg,Juer:1982fb,Juer:1982mp}, and dots denote the higher order terms.\footnote{The expressions for $\ln Z_c$ and $\ln Z_V$ are written for the case $y=0$.} Then (for $y=0$) the two-loop contributions to the anomalous dimensions of the quantum superfields in the $P=\frac{1}{3}Q$ theories can be written as

\begin{eqnarray}\label{Two_Loop_Gamma_C}
&& \gamma_c^{(2)} = \frac{\alpha^2}{\pi^2}\bigg\{\frac{C_2 Q}{12} - \frac{1}{24} (C_2)^2 \big(\xi^2-1\big) +\frac{1}{72} (C_2)^2\Big(4x_1 - (\xi-1) k_1\Big) -\frac{1}{4} (C_2)^2 \Big(\ln a_\varphi - b_{11}\Big)
\qquad\nonumber\\
&&\qquad + \frac{1}{12} C_2 T(R) \Big(\ln a -b_{12}\Big) + \frac{C_2 Q}{2} h_1 \bigg\};\\
&&\vphantom{1}\nonumber\\
\label{Two_Loop_Gamma_V}
&& \gamma_V^{(2)} = \frac{\alpha^2}{\pi^2}\bigg\{- \frac{Q^2}{24} - \frac{C_2 Q}{12} + \frac{1}{24}(C_2)^2  \big(\xi^2-1\big) - \frac{1}{72} (C_2)^2\Big(4x_1 - (\xi-1) k_1\Big) + \frac{1}{4} (C_2)^2 \Big(\ln a_\varphi
\nonumber\\
&&\qquad - b_{11}\Big)  -\frac{1}{12} C_2 T(R) \Big(\ln a - b_{12}\Big) + \frac{1}{8} Q\Big( -3 C_2 (b_{11} - v_1) + T(R) (b_{12} - v_2)\Big) \bigg\};\\
&& \vphantom{1}\nonumber\\
\label{Two_Loop_Gamma_Phi}
&& \big(\gamma^{(2)}_\phi\big)_i{}^j = -\frac{\alpha^2 Q^2}{36\pi^2} \delta_i^j + \bigg(\frac{\alpha^2 Q}{12\pi^2} C(R)_i{}^j +\frac{\alpha^2}{2\pi^2}\big[C(R)^2\big]_i{}^j + \frac{\alpha}{4\pi^3} \lambda^*_{imn} \lambda^{jml} C(R)_l{}^n\bigg) \Big(B-A \nonumber\\
&&\qquad  - 2 g_{11} + 2 g_{12}\Big) - \frac{3\alpha^2}{2\pi^2} C_2 C(R)_i{}^j \Big(\ln a_\varphi + \frac{1}{2}(1+A) - b_{11} + g_{11}\Big) + \frac{\alpha^2}{2\pi^2} T(R) C(R)_i{}^j\nonumber\\
&&\qquad \times \Big(\ln a + \frac{1}{2}(1+A) - b_{12} + g_{11}\Big),
\end{eqnarray}

\noindent
where

\begin{equation}
A \equiv \int\limits_0^\infty dx\,\ln x\, \frac{d}{dx}\Big(\frac{1}{R(x)}\Big);\qquad B \equiv \int\limits_0^\infty dx\,\ln x\, \frac{d}{dx}\Big(\frac{1}{F^2(x)}\Big);\qquad a\equiv \frac{M}{\Lambda};\qquad a_\varphi \equiv \frac{M_\varphi}{\Lambda}
\end{equation}

\noindent
are the regularization parameters. Here the functions $R(x)$ and $F(x)$ are the higher derivative regulators in the gauge and matter parts of the action, respectively, while $M$ and $M_\varphi$ are the masses of the Pauli--Villars superfields, see \cite{Aleshin:2016yvj,Kazantsev:2017fdc} for details. Note that the renormalization of the parameter $y$ is very essential for calculating the anomalous dimensions $\gamma_c$ (see \cite{Kazantsev:2018kjx}) and $\gamma_V$.

Substituting the expressions (\ref{Two_Loop_Gamma_C}) --- (\ref{Two_Loop_Gamma_Phi}) into Eq. (\ref{Gamma_Relation}) we see that in the two-loop approximation it is valid if the finite constants fixing a renormalization prescription satisfy the constraints

\begin{eqnarray}\label{Two_Loop_Scheme}
&& \ln a_\varphi + \frac{1}{2}(1+A) - b_{11} + g_{11} = 0;\qquad \ln a + \frac{1}{2}(1+A) - b_{12} + g_{11} = 0;\nonumber\\
&&  B-A - 2 g_{11} + 2 g_{12} = 0;\qquad\, h_1 -\frac{3}{4}\big(b_{11} - v_1\big) = 0; \qquad\, b_{12} - v_2 = 0.\qquad
\end{eqnarray}

\noindent
Therefore, there are subtraction schemes in which Eqs. (\ref{Gamma_Relation}) and (\ref{Beta_VCC}) (and, therefore, Eq. (\ref{Beta_Gamma_Relation})) are satisfied in the two-loop approximation. If the higher derivative regulator function in the gauge fixing term is the same as in the gauge part of the action, $K(x)=R(x)$, then in the Feynman gauge $\xi=1$, $y=0$ the values of the finite constants corresponding to the $\overline{\mbox{DR}}$ scheme are given by the expressions

\begin{equation}\label{DR_Bar_Constants}
\quad g_{11} = -\frac{1}{2}-\frac{A}{2}; \qquad g_{12} = -\frac{1}{2}-\frac{B}{2}; \qquad b_{11} =v_1 = \ln a_\varphi; \qquad b_{12} = v_2 = \ln a; \qquad  h_1=0.\quad
\end{equation}

\noindent
(Some of them were found in \cite{Kazantsev:2020kfl}, and the others were calculated similarly by comparing the relevant renormalized Green functions. In particular, the values of $v_1$ and $v_2$ were obtained with the help of the one-loop two-point Green function of the quantum gauge superfield derived in \cite{Kazantsev:2017fdc}.) Substituting the expressions (\ref{DR_Bar_Constants}) into Eq. (\ref{Two_Loop_Scheme}) we see that the relation (\ref{Gamma_Relation}) is valid in the $\overline{\mbox{DR}}$ scheme in the two-loop approximation (in the gauge $\xi=1$, $y=0$).

\vspace*{9mm}

\begin{figure}[h]
\begin{picture}(0,3)
\put(5,0){\includegraphics[scale=0.3]{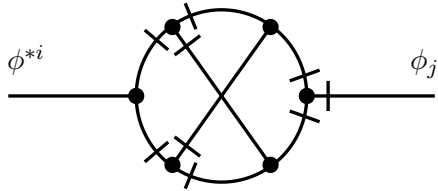}}
\put(5,1.6){$\phi^{*i}$} \put(10.3,1.6){$\phi_j$}
\end{picture}
\caption{The three-loop superdiagram contributing to the anomalous dimension of the matter superfields given by the expression (\ref{Yukawa_Combination}).}\label{Figure_Diagram}
\end{figure}

However, essential problems appear in the three-loop approximation. Really, the three-loop anomalous dimension of the matter superfields is contributed by the superdiagram presented in Fig. \ref{Figure_Diagram} given by the expression \cite{Parkes:1985hh}

\begin{equation}\label{Yukawa_Combination}
(\Delta\gamma_\phi)_i{}^j = \frac{3\zeta(3)}{64\pi^6}\, \lambda^*_{ikl} \lambda^{kpq} \lambda^{lrs} \lambda^*_{mpr} \lambda^*_{nqs} \lambda^{jmn},
\end{equation}

\noindent
and no other three-loop superdiagrams produce this combination of the Yukawa couplings. The contribution (\ref{Yukawa_Combination}) does not depend on a regularization and scheme parameters if a renormalization prescription satisfies Eq. (\ref{Lambda_Renormalization}). Such expressions can be calculated not only in the $\overline{\mbox{DR}}$ scheme, but even in the case of using the higher covariant derivative regularization, see, e.g., \cite{Shirokov:2022jyd}. In particular, in Appendix \ref{Appendix_Diagram} we derive the expression (\ref{Yukawa_Combination}) for the version of this regularization proposed in \cite{Aleshin:2016yvj,Kazantsev:2017fdc}. This calculation is useful because it explicitly demonstrates that the result is independent of regularization parameters. Evidently, it is impossible to simplify the expression (\ref{Yukawa_Combination}) with the help of the $P=\frac{1}{3}Q$ condition, so that Eq. (\ref{Gamma_Relation}) is not satisfied for all group structures if a renormalization prescription satisfies Eq. (\ref{Lambda_Renormalization}). In particular, it is not valid in the $\overline{\mbox{DR}}$ scheme in the three-loop approximation. Nevertheless, in principle, it is possible to absorb the structure (\ref{Yukawa_Combination}) into a finite redefinition of $\lambda^{ijk}$, but in the resulting scheme the renormalized Yukawa couplings do not satisfy Eq. (\ref{Lambda_Renormalization}). However \cite{Jack:1996qq}, with the help of similar finite renormalizations it is impossible to remove other terms proportional to $\zeta(3)$, and the anomalous dimension of the matter superfields cannot be made proportional to $\delta_i^j$. Therefore, to achieve the RG invariance of the ratio $\lambda^{ijk}/e$, one should either impose some additional constraints to the Yukawa couplings (certainly, if possible) or to suggest that Eq. (\ref{Gamma_Relation}) is valid only for a certain class of superdiagrams, e.g., for the planar supergraphs.

\section{Conclusion}
\hspace*{\parindent}

In this paper we demonstrated that the RG invariance of the ratio $\lambda^{ijk}/e$ in theories satisfying the $P=\frac{1}{3} Q$ constraint (\ref{PQ3_Constraint}) proposed in \cite{Jack:1995gm} leads to the equation (\ref{Gamma_Relation}). In any order of the perturbation theory this equation relates the anomalous dimensions of the quantum superfields (namely, of the Faddeev--Popov ghosts, of the quantum gauge superfield, and of the matter superfields). In the one-loop approximation this relation has been verified by an explicit calculation. For the verification of Eq. (\ref{Gamma_Relation}) in the next order we have constructed the expression for the two-loop anomalous dimension of the quantum gauge superfield using the finiteness of the triple gauge-ghost vertices, which has been verified explicitly in the two-loop approximation in \cite{Kuzmichev:2021yjo,Kuzmichev:2021lqa}. Then we obtained that the relation (\ref{Gamma_Relation}) is really valid in two loops for a certain class of renormalization prescriptions, which includes the $\overline{\mbox{DR}}$ scheme. However, in the three-loop approximation Eq. (\ref{Gamma_Relation}) is not valid because, according to \cite{Jack:1996qq}, there are no renormalization prescriptions in which the anomalous dimension of the matter superfields is proportional to $\delta_i^j$. Therefore, it is necessary either to impose some additional constraints on the Yukawa couplings or to suggest that the ratio $\lambda^{ijk}/e$ is RG invariant only in a certain approximation. However, if we assume that under certain conditions this is true, then the exact equation (\ref{Beta_Gamma_Exact}) for the anomalous dimension of the matter superfields can be rewritten in the equivalent form (\ref{Gamma_With_X}). The expression (\ref{Gamma_With_X}) does not contain the coupling dependent denominator and is analogous to the NSVZ $\beta$-function in the form (\ref{NSVZ_New}). It is this form that has been derived by direct perturbative summation of superdiagrams \cite{Stepanyantz:2020uke}, so that one expects that the derivation of Eq. (\ref{Beta_Gamma_Exact}) can possibly be made similarly using Eq. (\ref{Gamma_With_X}) with a certain value of $x$. Probably, this derivation could allow revealing the additional constraints which should be imposed on the Yukawa couplings or finding a class of superdiagrams for which Eq. (\ref{Gamma_Relation}) is valid. Taking into account that, according to \cite{Jack:1996qq}, all ``bad'' terms in the three-loop anomalous dimension of the matter superfields which survive in the $P=\frac{1}{3}Q$ theories and do not satisfy the second equation in Eq. (\ref{Beta_Gamma_Exact}) are proportional to $\zeta(3)$ one is tempted to suggest that this equation is valid for planar supergraphs for a certain renormalization prescription. However, this conjecture should be verified by explicit three-loop calculations which have not yet been done for an arbitrary subtraction scheme supplementing the higher covariant derivative regularization.

\section*{Acknowledgments}
\hspace*{\parindent}

The work of K.S. has been supported by Russian Science Foundation, grant No. 21-12-00129.

\appendix

\section*{Appendix}

\section{Derivation of Eq. (\ref{Yukawa_Combination}) in the case of using the higher covariant derivative regularization}
\hspace*{\parindent}\label{Appendix_Diagram}

After constructing the expression for the superdiagram presented in Fig. \ref{Figure_Diagram} we obtained that the corresponding contribution to the anomalous dimension (defined in terms of the bare couplings) is written as

\begin{equation}\label{Delta_Gamma}
(\Delta\gamma_\phi)_i{}^j = 16 \lambda^*_{0ikl} \lambda_0^{kpq} \lambda_0^{lrs} \lambda^*_{0mpr} \lambda^*_{0nqs} \lambda_0^{jmn} \cdot I,
\end{equation}

\noindent
where $I$ denotes the corresponding Euclidean loop integral

\begin{equation}\label{Loop_Integral}
I \equiv \frac{d}{d\ln\Lambda} \int \frac{d^4K}{(2\pi)^4} \frac{d^4L}{(2\pi)^4} \frac{d^4Q}{(2\pi)^4} \frac{1}{K^2 F_K L^2 F_L^2 Q^2 F_Q (K-Q)^2 F_{K-Q}^2 (L-Q)^2 F_{L-Q} (K-L)^2 F_{K-L}}.
\end{equation}

\noindent
Here Euclidean momenta are denoted by capital letters, $\Lambda$ is the dimensionful parameter of the higher covariant derivative regularization, and $F_K\equiv F(K^2/\Lambda^2)$ is the higher derivative regulator function present in the action of the chiral matter superfields (see \cite{Aleshin:2016yvj,Kazantsev:2017fdc} for more details.) Certainly, it is assumed that the derivative with respect to $\ln\Lambda$ is calculated before the integration. Then it is easy to see that the integral $I$ is a finite constant independent of $\Lambda$. To calculate it, first, we note that

\begin{equation}
\Big(\frac{d}{d\ln\Lambda} + K^\mu\frac{\partial}{\partial K^\mu} + L^\mu\frac{\partial}{\partial L^\mu} + Q^\mu\frac{\partial}{\partial Q^\mu}\Big) \frac{1}{F_K F_L^2 F_Q F_{K-Q}^2 F_{L-Q} F_{K-L}} = 0,
\end{equation}

\noindent
because the function $F$ is dimensionless. Using this equation and taking the limit $\Lambda\to \infty$ (in which $F_K\to 1$) we present the integral (\ref{Loop_Integral}) in the form

\begin{equation}\label{I_Expression}
I = 3\oint \frac{dS_\mu^{(Q)}}{(2\pi)^4} \int \frac{d^4K}{(2\pi)^4} \frac{d^4L}{(2\pi)^4} \frac{Q_\mu}{K^2 Q^2 L^2 (K-Q)^2 (L-Q)^2 (K-L)^2},
\end{equation}

\noindent
where the first integration is performed over the infinitely large sphere $S^3$ in the space with the Cartesian coordinates $Q_\mu$. The remaining integral can be calculated with the help of the Chebyshev polynomial method \cite{Rosner:1967zz} based on the equation

\begin{equation}\label{Inverse_Square_Momenta}
\frac{1}{(K-L)^2} = \left\{\begin{array}{l}
{\displaystyle \frac{1}{K^2} \sum\limits_{n=0}^\infty \Big(\frac{L}{K}\Big)^n C_n(\cos\theta),\quad\mbox{if}\quad K > L;}\\
\vphantom{1}\\
{\displaystyle \frac{1}{L^2} \sum\limits_{n=0}^\infty \Big(\frac{K}{L}\Big)^n C_n(\cos\theta),\quad\, \mbox{if}\quad L > K,}
\end{array}
\right.
\end{equation}

\noindent
where the Chebyshev polynomials are defined as

\begin{equation}
C_n(\cos\theta) \equiv \frac{\sin\, ((n+1)\theta)}{\sin\theta}.
\end{equation}

\noindent
Applying Eq. (\ref{Inverse_Square_Momenta}) and calculating the angular integrals in Eq. (\ref{I_Expression}) using the identities

\begin{eqnarray}\label{Product}
&& \int \frac{d\Omega_Q}{2\pi^2} C_m\Big(\frac{K_\mu Q^\mu}{KQ}\Big) C_n\Big(\frac{Q_\nu L^\nu}{QL}\Big) = \frac{1}{n+1} \delta_{mn} C_n\Big(\frac{K_\mu L^\mu}{KL}\Big);\\
\label{Ortogonality}
&& \int \frac{d\Omega}{2\pi^2} C_m(\cos \theta) C_n(\cos \theta) = \delta_{mn}
\end{eqnarray}

\noindent
we obtain

\begin{eqnarray}
&& I = \frac{6}{(8\pi^2)^3} \lim\limits_{Q\to \infty} Q^2 \sum\limits_{n=0}^\infty \frac{1}{(n+1)} \int\limits_0^Q dK\,K \int\limits_0^K dL\,L\, \frac{1}{Q^2} \Big(\frac{K}{Q}\Big)^n \frac{1}{Q^2} \Big(\frac{L}{Q}\Big)^n \frac{1}{K^2} \Big(\frac{L}{K}\Big)^n \qquad\nonumber\\
&& = \frac{3}{2(8\pi^2)^3} \sum\limits_{n=0}^\infty \frac{1}{(n+1)^3} = \frac{3}{2(8\pi^2)^3}\,\zeta(3).
\end{eqnarray}

\noindent
Substituting this expression into Eq. (\ref{Delta_Gamma}) the considered contribution to the anomalous dimension can be written as

\begin{equation}
(\Delta\gamma_\phi)_i{}^j = \frac{3\zeta(3)}{64\pi^6}\,\lambda^*_{0ikl} \lambda_0^{kpq} \lambda_0^{lrs} \lambda^*_{0mpr} \lambda^*_{0nqs} \lambda_0^{jmn}.
\end{equation}

\noindent
Taking into account that there are no contributions of the considered structure coming from the renormalization in the previous orders (in the subtraction schemes which satisfy Eq. (\ref{Lambda_Renormalization})), we see that the corresponding contribution to the anomalous dimension standardly defined in terms of the renormalized coupling can be obtained by the formal substitution $\lambda_0^{ijk} \to \lambda^{ijk}$.

\end{document}